\documentclass[preprint]{aastex}

\def\newpage{\vfill\eject} 
\newcommand{\be}{\begin{equation}}
\newcommand{\ee}{\end{equation}}
\newcommand{\mpl}{ {M_{\rm pl}}} 
 
\newcommand\pp{\parshape 2 0.0truecm 15.5truecm 1.25truecm 14.25truecm} 
\newcommand{\vac}{ \Omega_{\rm v,0} } 
\newcommand{\om}{ \Omega_{\rm m,0} }

\newcommand{\rwig}{ {\tilde r}} 

\begin{document}
\baselineskip=20pt 

\title{THE ASYMPTOTIC STRUCTURE OF SPACE-TIME
\footnote{This essay received ``Honorable Mention'' in the 
2003 Essay Competition of the Gravity Research Foundation} }

\author{\bf Fred C. Adams, Michael T. Busha, August E. Evrard, 
and Risa H. Wechsler}  

\affil{Michigan Center for Theoretical Physics, 
Univ. of Michigan, Ann Arbor, MI 48109} 

\email{fca@umich.edu} 

\begin{abstract} 

Astronomical observations strongly suggest that our universe is now
accelerating and contains a substantial admixture of dark vacuum
energy. Using numerical simulations to study this newly consolidated
cosmological model (with a constant density of dark energy), we show
that astronomical structures freeze out in the near future and that
the density profiles of dark matter halos approach the same general
form. Every dark matter halo grows asymptotically isolated and thereby
becomes the center of its own island universe. Each of these isolated
regions of space-time approaches a universal geometry and we calculate
the corresponding form of the space-time metric.

\end{abstract} 
 
\bigskip 
$\,$
\bigskip 

{\bf Introduction.} The basic cosmological parameters that describe
our universe have now been measured with compelling precision. Recent
measurements of the cosmic microwave background radiation indicate
that the universe is spatially flat [1]. Complementary measurements of
the redshift-distance relation using Type Ia supernovae strongly
suggest that the universe is now accelerating [2]. Taken together, the
current astronomical data argue for a cosmological model with matter
density $\om$ = 0.3, dark vacuum energy density $\vac$ = 0.7,
curvature constant $k=0$, and Hubble constant $H_0$ = 70 km s$^{-1}$
Mpc$^{-1}$. Although the time dependence of the dark energy has not
been fully determined, the current data are consistent with the vacuum
energy density being temporally constant, as this work assumes.

This newly consolidated cosmological model represents a milestone in
our understanding of the universe. The large scale space-time of the
universe is now known and its corresponding metric can be specified.
In the absence of structure formation, the universe is homogeneous and
isotropic, and the space-time would be described by the maximally
symmetric Robertson-Walker metric [3]. Since the universe does contain
gravitationally collapsed structures, however, the metric that
describes space-time is one step more complicated --- it must include
the contribution from the structures.

If the universe is already starting to accelerate, as observations
indicate, then structure formation is virtually finished. In the
relatively near future, the universe will approach a state of
exponential expansion and growing cosmological perturbations will
freeze out on all scales. Existing structures will grow isolated.
Because the parameters of our universe are now relatively well known,
this future evolution of cosmological structure can now be predicted
with a high degree of confidence. Several recent papers have begun to
explore the possible future effects of vacuum energy density [4--6],
and demonstrate that the universe will indeed break up into a
collection of ``island universes'', each containing one gravitational
bound structure.

In this essay, we present the results of a recent series of numerical
simulations that describe the evolution of structure in a universe
dominated by dark vacuum energy (with $\vac$ = 0.7 at the present
epoch). These numerical experiments show that each gravitationally
bound halo structure grows isolated and that its density profile
always approaches the same general form. After describing the
numerical simulations in greater detail and specifying the form of
this density profile, we construct the metric for each isolated patch
of space-time. Each island universe attains the same geometry and we
find the universal form for the metric that describes these patches of
space-time.

{\bf Numerical Simulations.} As part of a more comprehensive study of
structure formation in the future of an accelerating universe, we have
performed a series of numerical simulations [7]. This set of
cosmological simulations used the GADGET numerical package [8] and was
run on an Intel parallel cluster (at U. Michigan Center for Academic
Computing). The simulations were set up using a standard suite of
initial conditions starting at scale factor $a$ = 0.05 [9], and were
evolved forward into the future until the scale factor had grown to
$a$ = 100. The cosmology was chosen to have the standard parameters
described above, with $\om$ = 0.3, $\vac$ = 0.7, and $H_0$ = 70 km
s$^{-1}$ Mpc$^{-1}$.  All of the work reported here uses this choice
of cosmological parameters.

The simulations followed the evolution of a cubic, periodic region
with comoving linear size 366 Mpc. Only the evolution of the dark
matter was computed and we only obtain information about dark matter
halos at relatively large spatial scales. The numerical resolution was
set by using 128$^3$ dark matter particles, each with an effective
mass of 9.57 $\times 10^{11} M_\odot$. The force resolution had a
constant value of 285 kpc.  With this force and mass resolution, the
inner workings of the galaxy formation process are not well-resolved,
but larger scale structures --- the dark matter halos containing most
of the mass --- are well characterized.

Hundreds of dark matter halos form within the volume of the universe
studied by the simulations.  The evolution from high redshift to the
present follows the now-standard scenario.  Most of the structure in
the universe is already in place by the present epoch with $a=1$. As
the universe evolves into the future, the structures grow more defined
and more isolated. As the accelerating universe continues to expand,
bound structures separate rapidly from each other [5--7]. In the long
term, existing cosmic structures remain bound but grow isolated, as
illustrated by Figure 1.  A large cluster will become effectively
isolated in about 120 Gyr, whereas a smaller structure (like our Local
Group) will grow isolated in about 180 Gyr.  These structures will be
embedded within an accelerating universe with a constant horizon
scale, where the horizon distance $r_H$ is given by 
\be 
r_H = \chi^{-1} = {c \over H_0} {2 \over \pi} 
\Bigl( {15 \over \vac} \Bigr)^{1/2} 
\approx 12,600 \, \, {\rm Mpc} \, .  
\label{eq:horizon}
\ee 
This horizon distance $r_H$ is not the same as the particle horizon,
but rather is essentially the Hubble radius. The distance scale $r_H$
provides an effective ``boundary for microphysics'' within the much
larger space-time of the universe [3]. The acceleration of the
universe effectively divides our present-day space-time into many
smaller ``island universes''. For this discussion, we consider the
center of each dark matter halo to lie at the center of its own island
universe. As we show next, these dark matter halos develop density
profiles with a universal form in both time and mass (for our chosen
cosmology).

{\bf Generic Form for the Density Profile.} Numerical simulations
indicate that cosmic structures, from galaxies to clusters, tend to
develop the same basic form for the density profiles of their dark
matter halos [7,10]. As a result, every island universe will attain
the same geometry for its space-time. In order to estimate the
geometry of these space-times, we must first estimate the (nearly
universal) form for the density profile of the dark matter halos.

Using the results from our numerical simulations, we have constructed
a composite dark matter halo from the 50 largest halos produced by one
realization of the simulation. These 50 halos are normalized so that
the mean interior density has the same value at the spatial scale
$r_{200}$ (the radius at which the enclosed density is 200 times the
critical density). With this normalization, the individual dark matter
halos show relatively little dispersion (with a mean of about 35
percent) and hence the composite average is well defined. The profiles
are close to being spherically symmetric (this point is discussed in
Refs. [7,11]) so we consider density distributions that depend only on
radius. The composite profiles are shown in Figure 2 for varying
cosmological epochs, starting from the present (top curve) and
extending to $a$ = 100 (bottom curve). Notice how the density profiles
display the same characteristic form over a wide range of epochs, with
each subsequent profile being a stretched version of the previous one.
This fact that dark matter halos tend to approach a universal form has
been noted earlier [10], although the previous composite profiles were
more limited in spatial extent and did not match smoothly onto the
background universe.

The density profile at every cosmological epoch can be fit with a 
spherical density profile of the form 
\be
\rho(r) = {\rho_0 \over r/r_S [1 + (r/r_S)^p]^{3/2} } 
[1 + r/r_\infty]^{1 + 3p/2} \, .  
\label{eq:rhopro} 
\ee 
This profile describes the basic radial dependence of dark matter
halos in the inner regions and matches smoothly onto the background
density of the universe at large radii. Using the parameters $r_S$ =
0.50 $r_{200}$ and $p$ = 1.8, the above functional form provides a
good fit to the numerically determined density profiles for all
epochs.  In order to match the profile onto the background density of
the universe, the remaining parameter $r_\infty$ must scale according
to $r_\infty = r_{\infty(0)} a^{6/(3p+2)}$, where the present-day
value $r_{\infty(0)}$ = 4.7 $r_{200}$. The resulting fits to the
density profiles are shown as the dashed curves in Figure 2. This
relatively simple function (eq. [\ref{eq:rhopro}]) applies over a
factor of 10 in halo mass scale, and fits the numerically calculated
density profiles over nearly 5 decades in radial scale, 11 decades in
density, and a factor of 100 in the scale factor $a$. Over this range,
the RMS departure of the fitted functions (eq. [\ref{eq:rhopro}]) from
the composite averages is 0.13 in $\log_{10} \rho$ (which corresponds 
to differences of $\sim 35 \%$ in $\rho$). 

{\bf Asymptotic Form for the Metric.}  Using the specified form
(eq. [\ref{eq:rhopro}]) for the density profile, we can now determine
the line element $ds^2$ for the space-time within the horizon distance
$r_H$ [12].  The center of the coordinate system is taken to be at the
center of the cluster (or galaxy) and the mass distribution is assumed
to be spherically symmetric. We begin by writing the line element in
the form 
\be
ds^2 = - \Bigl(1 - A(r) - \chi^2 r^2 \Bigr) dt^2 + 
\Bigl(1 - B(r) - \chi^2 r^2 \Bigr)^{-1} dr^2 + r^2 d\Omega^2 \, , 
\label{eq:lineone} 
\ee
where we have explicitly separated out the the contribution due to the
cosmological constant, which is set by the parameter $\chi^2 \equiv 
(2 \pi^3/45)^{1/2} \Lambda^2/\mpl$ (where the energy scale $\Lambda 
\approx$ 0.0003 eV for $\vac$ = 0.7). In an ``empty'' universe
containing only vacuum energy, the line element would have the above
form with $A=0=B$.  Because of the vacuum contribution, the metric
contains an outer horizon at $r_H = \chi^{-1}$. This outer horizon
supports the emission of radiation through a Hawking-like mechanism
[13] and hence the future universe will be filled with a nearly thermal 
bath of radiation with temperature $T \sim \chi \sim 10^{-33}$ eV and
characteristic wavelength $\lambda \sim r_H \sim 12,600$ Mpc. This
radiation will become the dominant background radiation field after
about one trillion years. The functions $A(r)$ and $B(r)$ take into
account additional curvature due to the mass distribution, which has a
density profile given by equation [\ref{eq:rhopro}].

If we adopt units in which $c=1$ (and hence $G = \mpl^{-2}$), 
the function $B(r)$ can be written in the form 
\be
B(r) = 2G {m(r) \over r} = 8 \pi G {1 \over r} \int_0^r 
\rho(\rwig) \rwig^2 d\rwig \, , 
\ee 
where the density profile $\rho (r)$ is given by equation
[\ref{eq:rhopro}].  Since we are interested in the asymptotic form for
the metric, we can consider late times for which the scale $r_\infty$
is stretched beyond the horizon $r_H$. In this limit, the function
$B(r)$ can be simplified to the form 
\be 
B(r) = 4 \pi G \rho_0 r_S^2 {1 \over \xi} \int_0^\xi 
{x dx \over (1 + x^p)^{3/2} } \, \equiv \eta_0 \beta(\xi) \, . 
\ee 
In the second equality, we have defined the parameter $\eta_0 = 4 \pi
G \rho_0 r_S^2$ which sets the ``strength'' of the curvature and the
dimensionless function $\beta(\xi)$ which specifies the radial
dependence of the metric coefficient (where $\xi = r/r_S$). For
typical values, the strength parameter $\eta_0 \approx 10^{-6}$,
indicating that the departure from flatness is relatively small.
The resulting function $\beta(\xi)$ is shown in Figure 3. 

The function $A(r)$ is related to the usual gravitational potential
$\Phi$ through the definition ${\rm e}^{2 \Phi} \equiv 1-A(r)$ [12], 
where the potential is defined through the source equation 
\be
{d \Phi \over d r} = {G [m(r) + 4 \pi r^3 p] \over r (r - 2Gm) } \, .
\label{eq:source} 
\ee 
In this setting, the mass is dominated by collisionless dark matter 
particles and the pressure $p$ is negligible. Furthermore, the potential 
is small so that we can use the approximation ${\rm e}^{2 \Phi} \approx 
1 + 2 \Phi$ and hence $A(r) = - 2 \Phi (r)$, with $\Phi$ given by the 
integral of equation [\ref{eq:source}]. As a result, the function $A(r)$ 
can be written in the form 
\be
A(r) = A_\infty - \int_0^r {2 G m(r) \over r} 
{dr/r \over 1 - 2 G m(r)/r} = \eta_0 \Bigl[ \alpha_\infty - 
\int_0^\xi {\beta(\xi) \over 1 - \beta(\xi)} {d\xi \over \xi} 
\Bigl] \, \equiv \eta_0 \alpha(\xi) \, , 
\ee
where $\eta_0$ and $\beta(\xi)$ are as defined previously. We have
also defined an analogous dimensionless function such that $A(r)$ =
$\eta_0 \alpha(\xi)$. The quantity $A_\infty$ and its dimensionless
counterpart $\alpha_\infty$ are defined so that the potential $\Phi$
vanishes at spatial infinity [12]. As before, the dimensionless
parameter $\eta_0 = 4 \pi G \rho_0 r_S^2 \approx 10^{-6}$ sets the
level of the curvature.  The resulting function $\alpha(\xi)$ is shown
in Figure 3. This completes the specification of the metric.

{\bf Summary.} In this essay, we have constructed the asymptotic form
of the metric that describes space-time in our cosmological future.
Using numerical simulations, we have demonstrated that individual
gravitationally bound structures will become isolated in the near
future and thereby become their own ``island universes'' (Figure 1).
Each of these gravitationally bound entities --- dark matter halos ---
will attain a characteristic form for its density distribution (see
Figure 2 and equation [\ref{eq:rhopro}]). Finally, each bound
structure will live at the center of its own island universe, and the
metric of the surrounding space-time can be described by a line
element of the form 
\be
ds^2 = - \Bigl(1 - \eta_0 \, \alpha(\xi) - \chi^2 r^2 \Bigr) dt^2 + 
\Bigl(1 - \eta_0 \, \beta(\xi) - \chi^2 r^2 \Bigr)^{-1} dr^2 + 
r^2 d\Omega^2 \, , 
\label{eq:linefinal} 
\ee 
where $\eta_0 = 4 \pi G \rho_0 r_S^2$, $\xi = r / r_S$, and where
$\alpha(\xi)$ and $\beta(\xi)$ are shown in Figure 3. Astronomical 
entities (planets, stars, and galaxies) living within the universe
will continue to evolve over much longer time scales [4,14], but 
space-time itself can be described by equation [\ref{eq:linefinal}] 
for the vast majority of the total life of the universe.

The idea that some type of dark energy could affect the expansion of
the universe dates back to Einstein's original introduction of a
cosmological constant. Although this idea has been called Einstein's
greatest blunder, the currently observed cosmic acceleration suggests
that this concept may become one of Einstein's greatest legacies. The
motivation for the cosmological constant was to keep the cosmos static. 
In a twist of irony, the observed dark vacuum energy does not make the
universe static, but rather drives it to expand at an accelerating 
rate. But even though the universe expands and changes, and its
constituent astrophysical objects age, this essay shows that the local
space-time metric does approach a ``static'' asymptotic form 
(eq. [\ref{eq:linefinal}]). 

\newpage

{\bf References} 

\par\pp[1] 
D. N. Spergel et al., submitted to Astrophys. J., 
astro-ph/0302209 (2003). 

\par\pp[2] 
A. G. Riess et al., Astron. J. {\bf 116}, 1009 (1998); 
S. Perlmutter et al., Astrophys. J. {\bf 517}, 565 (1998); 
P. H. Garnavich et al. Astrophys. J. {\bf 507}, 74 (1998).  

\par\pp[3] 
E. W. Kolb and M. S. Turner, {\sl The Early Universe} 
(Addison-Wesley, Redwood City, 1990).   

\par\pp[4]
F. C. Adams and G. Laughlin, Rev. Mod. Phys. {\bf 69}, 337 (1997). 

\par\pp[5] 
A. Loeb, Phys. Rev. D {\bf 65}, 047301, (2002); 
K. Nagamine and A. Loeb, astro-ph/0204249 (2002).  

\par\pp[6]
E. H. Gudmundsson and G. Bj{\"o}rnsson, Astrophys. J. {\bf 565}, 1 (2002); 
T. Chieuh and X.-G. He, Phys. Rev. D {\bf 65}, 123518 (2002). 

\par\pp[7]
M. T. Busha, F. C. Adams, R. H. Wechsler, and A. E. Evrard, submitted 
to Astrophys. J. (2003). 

\par\pp[8]
V. Springel, N. Yoshida, and S.D.M. White, New Astronomy {\bf 6}, 
157 (2000).  

\par\pp[9] 
J. J. Bialek, A. E. Evrard, and J. J. Mohr, Astrophys. J. {\bf 555}, 
597 (2001). 

\par\pp[10]
J. F. Narravo, C. S. Frenk, and S.D.M. White, Astrophys. J. {\bf 490}, 
493 (1997). 

\par\pp[11]
Y. P. Jing and Y. Suto, Astrophys. J. {\bf 574}, 538 (2002). 

\par\pp[12]
C. W. Misner, K. S. Thorne, and J. A. Wheeler, 
{\sl Gravitation} (San Francisco: W. H. Freeman, 1973).  

\par\pp[13] 
N. D. Birrell and P.C.W. Davies, {\sl Quantum Fields in Curved 
Space-Time} (Cambridge, Cambridge Univ. Press, 1982). 
 
\par\pp[14]
F. J. Dyson, Rev. Mod. Phys. {\bf 51}, 447 (1979); 
J. N. Islam, Quart. J. Roy. Astron. Soc. {\bf 18}, 3 (1977);  
M. M. Cirkovic, Am. J. Phys. Resource Letters {\bf 71}, 122 (2003).

\newpage 
\begin{figure}
\figurenum{1}
{\centerline{{\epsscale{0.6} \plotone{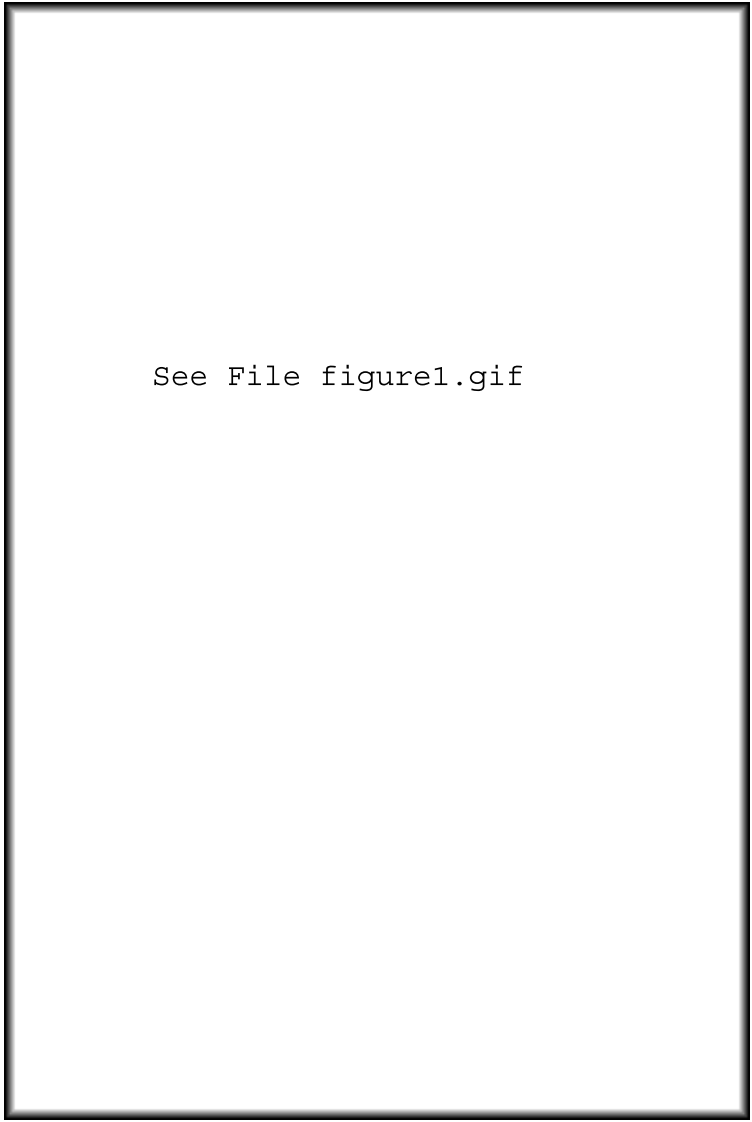} } }} 
\figcaption{Results of numerical simulations of structure formation 
in an accelerating universe with a constant density of dark vacuum
energy. Top panel shows a portion of the universe at the present epoch
when the scale factor $a$ = 1 (cosmic age 14 Gyr). The box in the
upper panel shows the region that expands to become the picture in the
center panel, which shows a portion of the universe at a future epoch
when $a$ = 11.4 (cosmic age 54 Gyr).  The box in the center panel
expands to become the picture shown in the bottom panel when the scale
factor $a$ = 100 (cosmic age 92 Gyr). By this future epoch, the dark
matter halo in the center of the bottom panel has grown effectively
isolated.}  
\end{figure}

\newpage 
\begin{figure}
\figurenum{2}
{\centerline{\epsscale{0.90} \plotone{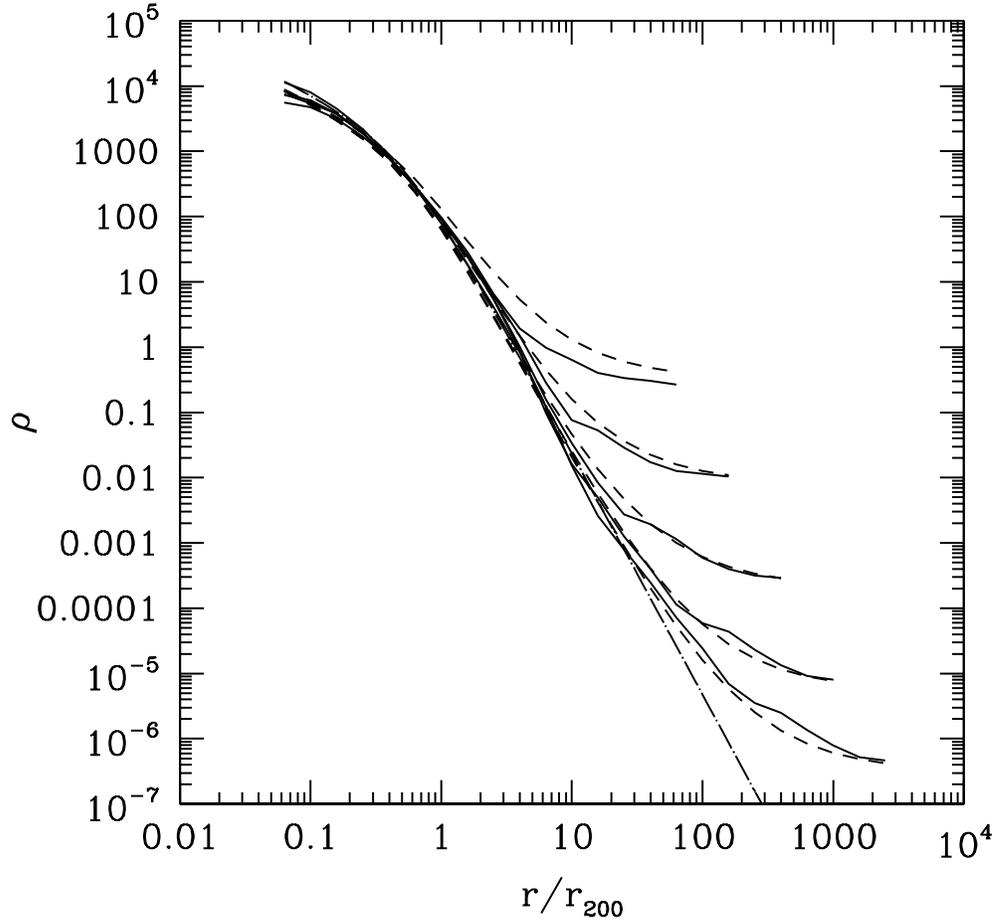} }}
\figcaption{The density profile for dark matter halos. Each curve
shows the average of the 50 largest dark matter halos in the numerical
simulation for a given time, ranging from the present epoch $a=1$ (top
curve) to $a=100$ (bottom curve). The numerically determined results
(averaged together) are shown as the solid curves.  The dashed curves
show the fits to the numerical results obtained from the analytic 
density profile of equation [\ref{eq:rhopro}].  The dot-dashed curve
shows the asymptotic form of the density profile (in the limit $t,a
\to \infty$).}  
\end{figure}

\newpage 
\begin{figure}
\figurenum{3}
{\centerline{\epsscale{0.90} \plotone{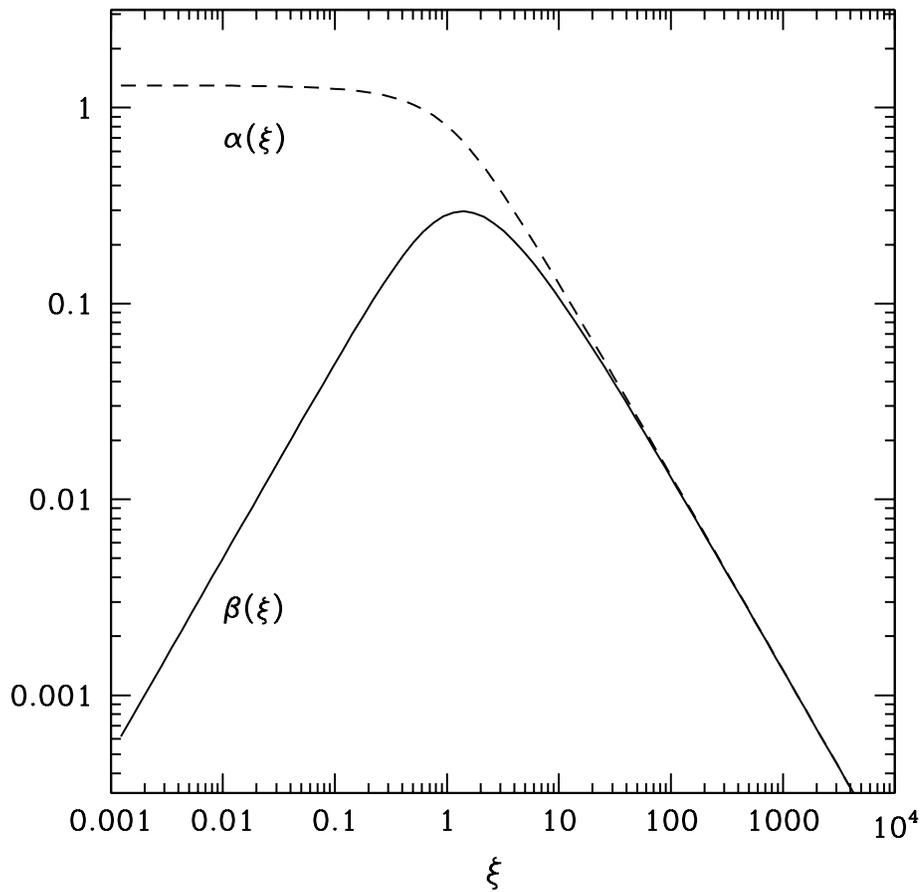} }}
\figcaption{The dimensionless functions $\alpha(\xi)$ and $\beta(\xi)$
appearing in the asymptotic form of the space-time metric of equation
[\ref{eq:linefinal}]. The functions are plotted versus the
dimensionless radial coordinate $\xi = r/r_S$ (see text). These
functions, in conjunction with equation [\ref{eq:linefinal}], specify 
the line element for the majority of the life of the universe. }
\end{figure}

\end{document}